\documentclass[a4paper, conference,10pt]{IEEEtran}
\usepackage{ifpdf}
\ifpdf
\else
 \fi
\usepackage{siunitx} % Required for alignment

\usepackage{cite}
\usepackage{tikz}

\ifCLASSINFOpdf
\else
\fi
\usepackage[cmex10]{amsmath}
\usepackage{amssymb}

\usepackage{algorithmic}
\usepackage{algorithm}
\renewcommand{\baselinestretch}{0.98}

\usepackage{array}
\usepackage[tight,footnotesize]{subfigure}
\usepackage[utf8]{inputenc}
\usepackage[english]{babel}
\usepackage{lettrine}

\usepackage{amsthm}
\usepackage{mathtools}
\usepackage{graphicx}
\usepackage{geometry}
\newcommand{\subparagraph}{}
\usepackage[compact]{titlesec}
\titlespacing{\section}{0pt}{1ex}{1ex}
\titlespacing{\subsection}{0pt}{1ex}{1ex}
\usepackage[font=small,skip=2pt]{caption}
\newtheorem{thm}{Theorem}
\newtheorem*{remark}{Remark}
\newtheorem{lemma}{Lemma}
\begin{document}
\title{Asymptotic Performance Analysis of Non-Bayesian Quickest Change Detection with an Energy Harvesting Sensor}

\author{Subhrakanti Dey \\
 Hamilton Institute, National Uni. of Ireland, Maynooth, Ireland \\ e-mail: Subhra.Dey@mu.ie }


%Telephone: (800) 555--1212, Fax: (888) 555--1212}


\maketitle

\begin{abstract}
In this paper, we consider a non-Bayesian sequential change detection  based on the Cumulative Sum (CUSUM) algorithm employed by an energy harvesting sensor where the distributions before and after the change are assumed to be known. In a slotted discrete-time model, the sensor, exclusively powered by randomly available harvested energy, obtains a sample and computes the log-likelihood ratio of the two distributions if it has enough energy to sense and process a sample. If it does not have enough energy in a given slot, it waits until it harvests enough energy to perform the task in a future time slot. We derive asymptotic expressions for the expected detection delay (when a change actually occurs), and the asymptotic tail distribution of the run-length to a false alarm (when a change never happens). We show that when the average harvested energy ($\bar H$) is greater than or equal to the energy required to sense and process a sample ($E_s$), standard existing asymptotic results for the CUSUM test apply since the energy storage level at the sensor is greater than 
$E_s$ after a  sufficiently long time. However, when the $\bar H < E_s$, the energy storage level can be modelled by a positive Harris recurrent Markov chain with a unique stationary distribution. Using asymptotic results from Markov random walk theory and associated nonlinear Markov renewal theory, we establish asymptotic expressions for the 
expected detection delay and asymptotic exponentiality of the tail distribution of the run-length to a false alarm in this non-trivial case. Numerical results are provided to support the theoretical results.
\end{abstract}
\IEEEpeerreviewmaketitle

\section{Introduction}
Sequential change point detection is an important task in many applications such as infrastructure safety monitoring, detection of sensor faults in unmanned autonomous vehicles, 
chemical process control, monitoring biological waster water treatment plants, intrusion detection in cyber-physical systems etc. \cite{ref_tar_book}. In general, sensors sequentially take samples of the monitored process and aims to detect a change in the statistical behaviour of the observe samples in the quickest possible fashion. Quickest change detection has been an active area of research for many decades \cite{ref15}. One of the optimal change detection method is given by the Cumulative SUM (CUSUM) method, where the change is to be detected as soon as possible after it happens by minimizing the supremum average detection delay subject to a constraint on the average run-length to a false alarm (when a change is detected even though no change has occurred. The CUSUM algorithm is based on a repeated application of a sequential probability ratio test (SPRT), where a sum of log-likelihood ratios between the distributions after and before the change, computed at the observed samples, is compared against a threshold, and a change is declared if the threshold is exceeded, and no change is declared otherwise. The threshold is chosen based on the average run-length to a false alarm constraint. Two of the most important performance measures related to any change detection method are (i) average run-length to detection, or expected detection delay (when a change has actually taken place), and (ii) average run-length to a false alarm. Various asymptotic expressions for expected detection delay and the tail distribution (in particular, asymptotic exponentiality) of the run-length to a false alarm have been shown in a number of works - see \cite{ref_tar_book} and references therein. 

In this paper, we consider a non-Bayesian quickest change detection in a slotted discrete-time scenario, where the observing sensor is solely powered by a random energy harvesting process. In such a scenario, when a sensor does not have enough available energy to sense and process a sample (denoted by $E_s$) of the observed phenomenon, the CUSUM test is temporarily halted, and it resumes again when the sensor has enough energy to obtain a sample and compute the log-likelihood ratio. Under the assumption of an independent and identically distributed harvested energy level in different time slots, we obtain asymptotic expressions for the average detection delay, and the asymptotic tail distribution of the run-length to a false alarm. In particular, we show that when the average harvested energy is greater than or equal to $E_s$, the energy storage level at the sensor will always be greater than 
$E_s$ asymptotically in time, and therefore after a sufficiently large amount of time 
(in practice, this may be only a short amount of time since $E_s$ is not expected to be excessive), the sensor will be able to take samples at every discrete-time slot, and therefore the standard asymptotic results regarding the expected detection delay and asymptotic exponentiality of the tail distribution of the run-length to a false alarm applies. In the case 
where the average harvested energy is less than $E_s$, we show that the underlying random walk in the modified CUSUM process is a Markov random where the energy storage level at the sensor asymptotically reaches a steady state distribution. Using a two-state Markov chain to define whether the energy storage process is greater or equal to $E_s$, or less than $E_s$, we show that this Markov chain is strongly recurrent, irreducible and aperiodic, where one can compute the steady state probabilities  of the two states numerically. 
Using asymptotic theory for first passage times and its tail distribution for a Markov random walk and associated nonlinear renewal theory \cite{lai_fuh_98, tartakovsky_fuh_HMM, 
karlin_dembo_92}, we prove similar asymptotic results for the expected detection delay and the asymptotic exponentiality of the tail distribution of the first passage time to a false alarm in this case. Note that while some earlier results regarding average detection delay for sequential detection with an energy harvesting sensor appeared in 
\cite{ref20, ref21}, these results were limited to a very simple Bernoulli arrival process for the harvested energy, whereas in the current work, we use a more general continuous-valued random process for the harvested energy. This general model significantly complicates the analysis (especially when the average harvested energy is less than $E_s$). Also, the asymptotic tail distribution results for the run-length to a false alarm have not appeared in the literature for the energy harvesting case to the best of the author's knowledge.

\section{Sequential Change Detection with Energy Harvesting}
In this section, we first provide some background theory on the traditional non-Bayesian quickest change detection problem where a sensor has no energy restrictions and can continuously sample a random process to perform a sequential probability ratio (SPRT) test. 
We then describe  how the sequential test is affected when the sensor is powered by harvested energy and is unable to sense and process a sample in case the energy storage at the sensor is less than the amount of energy required to sense and process at a given time. 

\subsection{Background on quickest change detection}
In this section, we focus on a non-Bayesian quickest change detection problem where a sensor observes a random process with {\em independent} discrete-time samples $\{X_k\}$, such that
\begin{align*}
X_k \sim \left\{ \begin{array}{ccc} F_0(x) & \text{if} & 1 \leq k \leq \nu \\    
                                                           F_1(x)  &   \text{if} & k \geq \nu+ 1  \end{array} \right.
\end{align*}
where $F_0, F_1$ are the cumulative distribution functions (c.d.f) before and after the change, with the corresponding probability density functions $f_0, f_1$, respectively. We assume that $f_1$ is absolutely continuous with respect to $f_0$. The change-point $\nu$ is unknown but deterministic. 

The objective of the quickest change detection problems is to detect the change-point $\nu$ as soon as possible after the change, if a change has occurred ($\nu < \infty$). Here we turn to Pollak's revised version of Lorden's formulation \cite{ref15,ref_tar_book}, where the following definitions are used. The {\em Supremum Average Detection Delay (SADD)} is defined to be 
\begin{align}
& SADD(T) = \sup_{0 \leq \nu < \infty} E_{\nu} (T - \nu | T > \nu), \end{align}
and the {\em Average Run Length (ARL) to False Alarm (ARL2FA)} is defined as $E_{\infty} T$ which denotes the average time to detect a change when the change never happens ($\nu = \infty$). The quickest change detection problem then can be formulated as 
\begin{align} 
& \text{Minimize} \:\: SADD(T)\: \:  \text{subject to} \:\:  E_{\infty} T \geq \gamma, \forall \gamma > 1, \end{align}
which is also known as the {\em minimax} formulation. It is well known that the {\em Cumulative Sum} (CUSUM) test (described below) is first-order asymptotically optimal for this minimax formulation \cite{ref16}. 

The CUSUM test is defined by the following test-statistic 
\begin{align}
& W_k = \max \left\{0, W_{k-1} + \log \frac{f_1(x_k)}{f_0(x_k)} \right\}, W_0 = 0  \label{cusum_orig} \end{align}
where $Z_k =  \log \frac{f_1(x_k)}{f_0(x_k)}$ is the log-likelihood ratio between the p.d.f after and before the change. 
Defining the stopping time $\tau(h) = \inf \{ n \geq h : W_n  > h\}$, when the threshold $h$ is chosen such that 
$E_{\infty} \tau(h) = \gamma$, the first order asymptotic optimality result states that 
\begin{align}
& SADD(\tau(h)) = \frac{\log \gamma}{{\cal I}_{KL}} \left(1 + o(1) \right), \; \text{as} \:\: \gamma \rightarrow \infty 
\label{cusum_asymp}
\end{align} 
where ${\cal I}_{KL} = \int \log \frac{f_1(x)}{f_0(x)} f_1(x) dx $ is the Kullback-Leibler divergence measure between the distributions after and before the change. 

\subsection{CUSUM test with an energy harvesting sensor}
In this subsection we consider a sensor that is equipped with an energy harvesting device, harvesting a random amount of energy $H_k \geq 0$ from ambient sources during the $k$-th time slot, and stores it in an energy storage device  (e.g. a supercapacitor) of {\em infinite capacity}\footnote{We can also extend the results to the case when the the capacity of the energy storage device is finite but much larger than the amount of energy required to sense and process a sample ($E_s$).}. Denoting the energy available at the sensor at time as $B_{k}$, we have the following standard model for the time-evolution of the energy storage device
\begin{align}
& B_{k+1} =  B_k + H_k - 1_{(B_k > E_s)} E_s, \label{batter_eq} \end{align}
where $E_s$ is the amount of energy required to sense a sample and process it in a sequential change detection algorithm, and $1_A$ is the indicator function taking value $1$ if and only if the event $A$ occurs, otherwise taking value $0$.  We have also made the assumption that the energy harvested during time-slot $k$ is only available for consumption at time-slot $k+1$. Clearly, if the sensor has less than $E_s$ amount of energy at the beginning of the $k$-th slot, it is unable to sense and process the sample $X_k$. This leads us to the following modified version of the CUSUM test 
\begin{align}
& \bar W_k = \max \left\{0, \bar W_{k-1} +\xi_k  \log \frac{f_1(x_k)}{f_0(x_k)} \right\}, \bar W_0 = 0,  \label{cusum_mod} \end{align}
where $\xi_k = 1_{(B_k > E_s)}$. Clearly, $\xi_k =1$ with $P(B_k \geq E_s)$ and $\xi_k = 0$ with $1 - P(B_k > E_s)$. The harvested energy process $\{H_k\}$ is assumed to independent and identically distributed  ({\em i.i.d.}) with an absolutely continuous (with respect to the Lebesgue measure) distribution  having a  finite mean $E(H_k) = \bar H$, and also independent of the sensed process
 $\{X_k\}$. 

In the next section, we show that the random process $\xi_k$ can be characterized according to the two possible scenarios: (i) 
$ \bar H \geq  E_s$ and (ii) $\bar H < E_s$.  For each of these  scenarios, we can analyze the performance of the modified CUSUM algorithm \eqref{cusum_mod} in terms of the average detection delay and the asymptotic distribution of the false alarm probability as $\gamma \rightarrow \infty$, the two most important performance metrics in the context of a sequential change point detection problem.

\section{Performance Analysis when $\bar H \geq E_s$}
In this section, we analyse the case when $\bar H \geq E_s$, and show that in this case, $P(\xi_k =1) = 1$ for a sufficiently large 
$k > N$. 
This result follows from the Strong  Law of Large Numbers (SLLN), when applied to the {\em i.i.d.} sequence $\{H_k\}$. 
Note that from \eqref{batter_eq}, we have
\begin{align*}
& B_k = B_0 + \sum_{l=0}^{k-1} H_l - E_s \sum_{l=0}^{k-1} 1_{(B_l > E_s)}
\end{align*}
Therefore the event $B_k > E_s$ holds iff 
\begin{align*}
& \frac{ \sum_{l=0}^{k-1} H_l}{k} > E_s \frac{\sum_{l=0}^{k-1} 1_{B_l > E_s}}{k} - \frac{B_0 -E_s}{k}
\end{align*}
We know from SLLN that there exists a sufficiently large $N(\epsilon)$, such that for $k > N(\epsilon)$, we have 
$\frac{ \sum_{l=0}^{k-1} H_l}{k} > \bar H - \epsilon$, for any $\epsilon > 0$. We can see that when $B_0 > E_s$,
\begin{align}
& E_s \frac{\sum_{l=0}^{k-1} 1_{B_l > E_s}}{k} - \frac{B_0 -E_s}{k} \leq E_s - \frac{B_0 -E_s}{k}  \nonumber \\
& \leq \bar H - \frac{B_0 -E_s}{k}  < \bar H - \epsilon < \frac{ \sum_{l=0}^{k-1} H_l}{k} 
\label{proof_SLLN}
\end{align}
where in the second last step, $\epsilon < \frac{B_0 -E_s}{k}$. When $B_0 \leq E_s$, one can easily modify the above proof and choose 
$\epsilon < \frac{B_0}{k}$ to satisfy the final inequality in \eqref{proof_SLLN}. 

Since $P(\xi_k =1) = 1$ for a sufficiently large 
$k > N$, as we are interested in the asymptotic scenario as $\gamma \rightarrow \infty$, \eqref{cusum_mod} reverts back to 
\eqref{cusum_orig} and we can apply existing results for the standard CUSUM test, as detailed in \cite{ref_tar_book}. 

We summarize the results for the average detection delay and the asymptotic distribution of the first passage time to a false alarm (FA) 
for this case in the next two subsections.

\subsection{Average Detection Delay}
In order to proceed, we define the following random walk $S_n= \sum_{k=0}^n Z_k, \; S_0 =0$, where $Z_k = \log \frac{f_1(x_k)}{f_0(x_k)}$, 
as defined earlier. Denoting the expectation under $f_1$ by $E_1$ (conditioned on the assumption that the change-point $\nu=1$), we have $E_1(Z_k) = {\cal I}_{KL}$. Define $E_1[(Z_k-{\cal I}_{KL})^2] = \sigma_1^2 < \infty$. Similarly, define the probability measure under $f_1$ as $P_1$. Define also the running minimum 
$\zeta_n = - \min_{0 \leq k \leq n} S_k$. Then it can be shown that $W_n$ from \eqref{cusum_orig} can be written as 
$W_n = S_n -  \min_{0 \leq k \leq n} S_k = S_n + \zeta_n$. Thus, $W_n$ appears as a perturbed version of the original random walk $S_n$. 

While the majority of the results regarding the first passage time for the random walk to reach a certain threshold were developed from the original random walk $S_n$, nonlinear renewal theory has made it possible to extend these results to the perturbed random walk $W_n$, provided the perturbation terms $\zeta_n$ satisfy the following ``slowly varying'' conditions: (i) $\frac{1}{n} \max_{1 \leq k \leq n} |\zeta_k| 
\rightarrow 0$, as $ n \rightarrow \infty$ (in probability), and (ii) for every $\epsilon > 0$, there are $N^* \geq 1$, and $\delta > 0$ such that 
$P\left(\max_{1\leq k \leq n \delta} |\zeta_{n+k } - \zeta_n | > \epsilon \right) , \epsilon, \forall n \geq N^*$. 
It has been shown that in case of the CUSUM algorithm \eqref{cusum_orig}, $\zeta_n = - \min_{0 \leq k \leq n} S_k$ satisfies these 
conditions - see p. 50 of \cite{ref_tar_book}. 

We recall the definition of the first passage time $\tau(h) = \inf \{ n \geq h : W_n  > h\}$, and define the overshoot 
$\kappa(h) = W_{\tau(h)} - h$.  Define also the first ladder epoch $T_{+} = \inf_{ n \geq 1 : S_n > 0 }$ and the corresponding ladder height 
$S_{T_{+}}$. Denote $X^{-} = - \min(0, X)$. 
Under the above mentioned ``slowly varying'' conditions, it has been shown that the asymptotic properties of the first passage time of  a standard random walk $S_n$ (where the underlying distribution $f_1$  is non-arithmetic, and has a positive mean and finite variance) extend to those of the perturbed random walk $W_n$. In particular, the following results hold from nonlinear renewal theory \cite{ref_tar_book} 
\begin{align}
& \lim_{h \rightarrow \infty} E_1 [ \kappa(h) ] = \kappa_{\infty} = \frac{E_1 [ S_{T_{+}}^2]}{2 E_1 [ S_{T_{+}}]} \nonumber \\ 
& = \frac{E_1[Z_1^2]}{2E_1[Z_1]} + E\left[ \min_{n \geq 0} S_n \right], \nonumber \\
& P_1(\bar \tau(h) \leq x, \kappa(h) \leq y) = \Phi(x) H(y), \nonumber \\ 
\:\:\: &  \text{as} \; h \rightarrow \infty, \: \forall x \in \{-\infty, \infty\},\:  y \geq 0
\label{non_renew_results}
\end{align}
where $\bar \tau(h) = \frac{\tau(h) - \frac{h}{{\cal I}_{LK}}}{\frac{h \sigma_1^2}{{\cal I}_{LK}^3}}$, and $\lim_{h \rightarrow \infty} P_1(\kappa(h) \leq y) = H(y)$, and $\Phi(x)$ is the c.d.f of the standard Normal distribution ${\cal N} (0, 1)$. 
Essentially, the first result above provides an accurate approximation for computing $E_1[\tau(h)] = \frac{1}{{\cal I}_{LK}}\left(h + E_1 [ \kappa(h) ] - E_1 [\zeta_{\tau(h)}]\right)$, which can be approximated as 
\begin{align*}
& E_1[\tau(h)] = \frac{1}{{\cal I}_{LK}}\left(h +\frac{E_1 [ S_{T_{+}}^2]}{2 E_1 [ S_{T_{+}}]}  - \bar \zeta \right) + o(1), \; \text{as} \; h \rightarrow \infty  \end{align*}
where it can be shown that $E_1[\zeta_n] \rightarrow \bar \zeta$, as $n \rightarrow \infty$. 

The second result in \eqref{non_renew_results} illustrates that the normalized first passage time $\bar \tau(h)$ and the overshoot asymptotically become independent as 
$h \rightarrow \infty$ and $\bar \tau(h)$ assumes a standard normal distribution asymptotically. While  we focus on the average detection delay in this paper, the asymptotic distribution is of importance when one has a distributed change detection scenario where multiple  sensors observe the change and make local decisions and send these to a fusion centre for making a decision using some fusion logic, such as based on the minimum/maximum of the first passage times of all sensors, or based on a majority vote from all the sensors etc. In this distributed case, computing the distribution of the minimum, maximum or median of the asymptotic distributions will provide a way to approximate the average detection delay, which will be investigated in a separate work.

Finally, noting that (see equation (8.152) in \cite{ref_tar_book} and see also the first result in \eqref{non_renew_results}) 
$\bar \zeta =  \frac{E_1[Z_1^2]}{2{\cal I}_{LK}}  - \kappa_{\infty}$, we can establish the following result for the average detection delay:
\begin{thm}
For an energy harvesting sensor employing  a CUSUM test \eqref{cusum_orig} to detect a change from $f_0$ to $f_1$ in the observed random variable, with an average harvested energy $\bar H \geq E_s$, the average detection delay under the alternative hypothesis ($f_1$) is independent of $\bar H$, and can be computed according to the following first order asymptotic approximation (as the detection threshold $h \rightarrow \infty$):
\begin{align}
E_1[\tau(h)] = \frac{1}{{\cal I}_{LK}}\left(h +\frac{E_1 [ S_{T_{+}}^2]}{ E_1 [ S_{T_{+}}]}  -  \frac{E_1[Z_1^2]}{2{\cal I}_{KL}} \right) + o(1)
\label{eq:LD_av_delay}
\end{align}
where recall that $Z_k = \log\frac{f_1(x_k)}{f_0(x_k)}$, and we have implicitly assumed that $\sigma_1^2 = E[(Z_k - {\cal I}_{KL})^2] < \infty$.
\label{LD_av_delay}
\end{thm}
\subsection{Asymptotic Distribution of the First Passage Time to a False Alarm}
For the purpose of this section, we need to consider a random walk $S_n = \sum_{i=1}^n  \bar Z_i, \; S_0= 0$ where $\bar Z_i$ is i.i.d. with a non-arithmetic distribution of mean $\bar \mu < 0$. Define the moment generating function $M(\theta) = E \exp (\theta \bar Z_1)$. It can be shown that there exists a unique $\gamma > 0$ such that 
$M(\gamma) = 1$. Define $\mu_{\gamma} = E[\bar Z_1 e^{\gamma \bar Z_1}] < \infty$. Then, with the associated reflected random walk 
$ W_k = \max \left(0, W_{k-1} + \bar Z_k\right)$, we define the first passage time $\bar \tau(h) = \inf\{ n \geq 1: W_n \geq h\}, \; h > 0$, and the first descending ladder epoch 
$T_{-} = \inf \{n \geq 1: S_n \leq 0\}$.  Then the following result has been proved in \cite{seq_det_khan}:
\begin{align}
& \lim_{ h \rightarrow \infty} P \left(e^{-\gamma h} \bar \tau(h) > x \right) = e^{-\beta x}, \; x > 0  \label{khan_result}
\end{align}
where $\beta = \left( \frac{1 - E e^{\gamma S_{T_{-}}}}{E[T_{-}]} \right)^2/\gamma \mu_{\gamma}$.

Essentially the above result states that the first passage time (appropriately scaled) for a random walk with a negative drift has asymptotically exponential tail as the threshold goes to infinity. It is not difficult to see the relevance of this result towards analyzing the average run length to false alarm of the CUSUM algorithm \eqref{cusum_orig} under the null hypothesis ($f_0$), where the increment is also i.i.d. with mean $-I_0 = - \int \log\left(\frac{f_1(x)}{f_0(x)}\right) f_0(x) dx$. Specializing to this case where the increments are log-likelihood functions given by $\bar Z_k = \log\frac{f_1(x_k)}{f_0(x_k)}$, it is obvious that $\gamma = 1$, since $E_{\infty}[e^{\gamma \bar Z_1}] = 1$ for $\gamma=1$, where 
$E_{\infty}$ denotes the expectation under the null hypothesis (i.e, the change never happens). Finally, in this case $\mu_{\gamma} = \int \log\left(\frac{f_1(x)}{f_0(x)}\right) 
\frac{f_1(x)}{f_0(x)}f_0(x) dx = \int \log\left(\frac{f_1(x)}{f_0(x)}\right) f_1(x) dx = {\cal I}_{LK}$. Now, under the null hypothesis, define $\tau_{\infty}(h) = \inf \{n \geq 1: W_n \geq h\}$, where $W_n$ is defined by \eqref{cusum_orig}. 

Using the above simplifications, and further renewal theoretic results from \cite{siegmund}, it was shown in \cite{pollak_tar_09} that the exponent $\beta$ in \eqref{khan_result}
can also be expressed as ${\cal I}_{KL} \bar \delta^2$, where $\bar \delta = \lim_{h \rightarrow \infty} E_1 \left[ \exp\{-(S_{\tau_{\infty}(h)} - h)\}\right]$, a renewal theoretic quantity that can be computed numerically. It should be noted that in \cite{pollak_tar_09}, the authors established the asymptotic exponentiality of the tail distribution of the first passage time to a false alarm for  more general Markov processes under suitable conditions.

Summarizing the above results, one can state the following theorem:
\begin{thm}
For an energy harvesting sensor employing  a CUSUM test \eqref{cusum_orig} to detect a change from $f_0$ to $f_1$ in the observed random variable, with an average harvested energy $\bar H \geq E_s$, the asymptotic tail distribution of the (normalized) first passage time to a false alarm is independent of $\bar H$, and is given by 
\begin{align}
&  P_{\infty} (e^{-h} \tau_{\infty}(h) > x) = e^{-\bar \beta x}, \; h \rightarrow \infty 
\end{align}
where {\small $\bar \beta = {\cal I}_{KL} \bar \delta^2, \; \bar \delta = \lim_{h \rightarrow \infty} E_1 \left[ \exp\{-(S_{\tau_{\infty}(h)} - h)\}\right]$}, and 
$E_{\infty}[\tau_{\infty}(h)] = \frac{e^h}{{\cal I}_{KL} \bar \delta^2}\left(1+o(1)\right)$. 

\label{tail_dist_FA_iid}
\end{thm}

\section{Performance Analysis when $\bar H  < E_s$}
In this section, we investigate the scenario when $\bar H < E_s$, and in the author's opinion, this turns out to be a more interesting scenario, although in practice, assuming $E_s$ is sufficiently small, we may be able to avoid this scenario. However, in multisensor distributed detection schemes, it may be true that a few sensors may not have favourable harvesting conditions and can fall into this category.  We show that in this case, the CUSUM statistic in \eqref{cusum_mod} can be described as a {\em reflected Markov Random Walk}.

\subsection{Stationarity of the battery state process $B_k$}
We first analyze the evolution of the battery state $B_k$ in the scenario and show that it is positive Harris recurrent Markov process with a unique invariant probability measure, or a stationary distribution. Although it is easier to prove such results in the case of a finite-discrete state space Markov chain, the proof is a little more complicated in the case where $B_k$ belongs to a general Borel state space. Wr first note that 
$B_{k+1} = (B_k - 1_{B_k > E_s}E_s) + H_k$, which implies that it is a nonlinear state space model. Since the distribution of  $H_k$ is continuous, and the Markov process $B_k$ satisfies the so-called ``forward-accessibility'' model (similar to controllability for linear systems, implying that for every given initial state, the set of all states reachable at some point in future is non-empty. Then, it follows from Proposition 7.1.2 in \cite{meyn_tweedie_book}, the Markov process $B_k$ is a {\em T-chain}, which is a slightly weaker property than a strong Feller chain \cite{meyn_tweedie_book}. It follows also that $B_k$ is a strong Feller chain and contains one reachable point, and therefore is irreducible (or more  technically, $\psi$-irreducible, 
see Proposition 6.1.5  \cite{meyn_tweedie_book}). Finally, consider the compact set ${\cal B}_s \coloneqq [0, E_s]$. Since $B_k$ is an irreducible T-chain, the set ${\cal B}_s $ is a {\em 
petite set} - see Proposition 6.2.5 of \cite{meyn_tweedie_book}.  

The above discussion allows us to apply the well known {\em Foster-Lyapunov} stochastic stability criterion for positive recurrence \cite{meyn_tweedie_book}. 
First note that one can rewrite \eqref{batter_eq} as 
$B_{k+1} = B_k + (H_k -E_s) + E_s 1_{(B_k \leq E_s)}$. Defining a Lyapunov function $V(B) = B$, we can see that 
$$ E[V(B_{k+1}) - V(B_k) | B_k =x] = \bar H - E_s + E_s 1_{(x \in {\cal B}_s)}. $$
Thus it follows that $ E[V(B_{k+1}) - V(B_k) | B_k =x] \leq - \bar \epsilon < 0$ (when $x \notin {\cal B}_s$), where $\bar \epsilon < \bar H - E_s$, and 
$ E[V(B_{k+1}) - V(B_k) | B_k =x] = \bar H < \infty$ when $x \in {\cal B}_s$. Therefore, from the {\em Foster-Lyapunov} stochastic stability criterion on  a $\psi$-irreducible Markov chain with a petite set ${\cal B}_s $, it follows that $B_k$ is a positive Harris  recurrent and has a unique invariant (stationary) measure. 

The above fact easily leads to the fact that the process $\xi_k \coloneqq 1_{(B_k > E_s)}$ is an aperiodic irreducible finite state Markov chain, thus having a unique stationary distribution. 
Note that while proving the existence of a stationary measure for the discrete Markov chain $\xi_k$ directly might have been straightforward, we wanted to establish the result for the general state space process $B_k$, so that it allows to compute the stationary distribution of $B_k$, and hence the transition probability distributions of $\xi_k$, namely 
$\tilde \beta = P(\xi_{k+1} = 1 | \xi_k =1)$ and $\tilde \alpha = P(\xi_{k+1} = 0 | \xi_{k} =0)$. We discuss the computation of this stationary distribution in the next subsection. 

\subsection{Computation of the stationary distribution of  $B_k$}
In the previous section, we established the existence and uniqueness of a stationary distribution of the Markov process $B_k$. Here we provide an integral equation that can be used to compute the stationary distribution numerically, given a continuous distribution $f_H(h)$ of the i.i.d. energy harvesting process $H_k$. Denoting the stationary distribution of the battery state $B_k$ by $f_B(b), \; b \geq 0$, the following Lemma can be derived:
\begin{lemma}
The stationary density $f_B(b)$ of the battery state (when $\bar H < E_s$) satisfies the following linear integral equation:
\begin{align}
& f_B(z) = \int_{E_s}^{z+E_s} f_H(z+E_s - b) f_B(b) db  \nonumber \\ 
& \quad \quad + \quad \int_{0}^{\min (z, E_s)} f_H(z-b) f_B(b) db, z \geq 0 \label{eq:stat_den_battery}
\end{align}
\label{battery_density}
\end{lemma}
Lemma \ref{battery_density} can be obviously used to compute the transition probabilities $\tilde \alpha,  \: \tilde \beta$ of the Markov chain $\xi_k$ as follows:
\begin{align*}
& \tilde \alpha = \frac{\int_0^{E_s} F_H(E_s -b) f_B(b) db}{\int_0^{E_s} f_b(b) db} \\
& \tilde \beta = \frac{ \int_{E_s}^{2E_s} \left(1-F_H(2E_s -b)\right) f_B(b) db + \int_{2E_s}^{\infty} f(b) db}{\int_{E_s}^{\infty} f_B(b) db},
 \end{align*}
where $F_H(h) $ is the c.d.f of the harvested energy process $H_k$. 

With the above analysis, we have now established that the process $\xi_k$ in \eqref{cusum_mod} is an aperiodic irreducible Markov chain with a transition probability matrix 
$\begin{bmatrix}
\tilde \alpha & 1- \tilde \alpha \\
1 - \tilde \beta &  \beta
\end{bmatrix}, \quad $ where the first row corresponds to the state $\xi_k =0$ and the second row corresponds to $\xi_k =1$, with the corresponding stationary distribution 
$\left[\frac{1-\tilde \beta}{(1-\tilde \alpha)+(1-\tilde \beta)} \; \frac{1-\tilde \alpha}{(1-\tilde \alpha)+(1-\tilde \beta)}\right]^{T}$. 

This leads to the crucial conclusion that in the case $\bar H < E_s$, \eqref{cusum_mod} is actually a special class of a {\em reflected Markov Random Walk} where the increment 
$\tilde Z_k \coloneqq \xi_k Z_k$ 
depends only on the current state of the Markov chain, and conditioned on the state trajectory of the Markov chain, the increments are i.i.d. Note that for a general Markov random walk, the increments may depend on both the current and past states of the associated Markov chain $(\xi_k, \: \xi_{k-1})$. 
\subsection{Average Detection Delay}
Once again, in order to proceed, we define the Markov random walk (MRW) $\tilde S_n = \sum_{k=0}^n \xi_k Z_k$, $\tilde S_0 =0$, where the underlying two-state Markov chain $\xi_k$ is aperiodic and irreducible with a unique stationary distribution $\pi = [\pi_0 \: \pi_1] = \left[\frac{1-\tilde \beta}{(1-\tilde \alpha)+(1-\tilde \beta)} \; \frac{1-\tilde \alpha}{(1-\tilde \alpha)+(1-\tilde \beta)}\right]$. 
Note also that the MRW only increments when $\xi_k=1$, otherwise remains static. This simplifying observation implies that $\tilde S_n  = \sum_{k=0}^n \tilde S_{t_k}$, 
with $\xi_{t_0}=1, S_{t_0} =0$, and $0 = t_0 < t_1 < \ldots < t_n \leq n$.  Therefore, it is apparent that the MRW at hand is also a sum of i.i.d. random variables $Z_{t_k}$, albeit with the random time instants $\{t_k\}$ being the sequence of time instants where the Markov chain $\xi_k$ visits state $1$.  Note that here we assume that the MRW is initialized at battery state $\xi_0 =1$, which is justified for two reasons: (i) we can always ensure that the battery has enough energy $(\geq E_s)$ to start with, and (ii) since the Markov chain is ergodic, even if it was intialized at state $0$, it will eventually visit state $1$ in finite time with nonzero probability, and this time to the first visit of state $1$ would not make a difference in the asymptotic case when $h \rightarrow \infty$. Therefore, in what follows, we will assume that the chain $\xi_k$ starts at $\xi_0 =1$.  We will denote the probability measure and expectations under the alternative hypothesis with subscript $1$ for the rest of this subsection. Finally, note that 
the mean of the MRW under the stationary distribution is given by $E_{\pi} [\xi_k Z_k] = \pi_1 E_1[Z_1] = \pi_1 {\cal I}_{KL}$. 

We now define the first passage time for the reflected Markov Random Walk \eqref{cusum_mod} as 
$\hat  \tau(h) = \inf \{ n \geq h : \bar W_n  > h\}$, and the corresponding first passage time for the associated MRW 
$\tilde \tau(h) =  \inf \{ n \geq h : \tilde S_n  > h\}$. While a sophisticated analysis of the expected first passage time, $E_1[\tilde \tau(h)]$ under certain finite moment assumptions  has been carried out in \cite{lai_fuh_98} (see Theorem 4) as $h \rightarrow \infty$, we actually need to obtain similar results for the first passage time 
$\hat  \tau(h)$. One would then expect that a similar nonlinear renewal theory for a Markov random walk can be applied by defining 
$\bar W_n = \tilde S_n -  \min_{0 \leq k \leq n} \tilde S_k = S_n + \eta_n$, where $\eta_n = -\min_{0 \leq k \leq n} \tilde S_k = -\min_{0 \leq t_k \leq n} \tilde S_{t_k}$, a ``slowly varying'' perturbation term. Indeed, such a nonlinear renewal theory for MRW can be found in a number of works, out of which we choose to follow \cite{tartakovsky_fuh_HMM} for its simplicity and relevance to our scenario. In particular, we refer the readers to Appendix A of \cite{tartakovsky_fuh_HMM}, which provides a synopsis of the analysis that we require. 

{\em Assumptions}: Note that in \cite{tartakovsky_fuh_HMM}, the asymptotic analysis is presented for a general state space Markov chain satisfying a concept of {\em V-uniform ergodicity}. However, for the scenario considered here, since the underlying Markov chain $\xi_k$ is  two-state, irreducible and aperiodic, with the assumption that 
$E[(Z_k - {\cal I}_{KL})^2] < \infty$ (as in Theorem 1), the additional assumptions required in order to obtain an asymptotic expression for the expected first passage time $\hat \tau(h)$ simplify to the following: (i) $\left\{\max_{1 \leq j \leq n} |\eta_{n+j}|,  n \geq 1 \right\}$ is uniformly integrable, (ii) $n P_1\{\max_{1 \leq j \leq n} \eta_{n+j} \geq \theta n\} \rightarrow 0$, as 
$n \rightarrow \infty$ for all $\theta > 0$, (iii) $\sum_{n=1}^{\infty} P_1(\eta_n \leq -\omega n) < \infty$ for some $0  < \omega < \pi_1 {\cal I}_{KL}$, and 
(iv) there exists $0 < \epsilon < 1$ such that 
$P_1(\bar  \tau(h) \leq \frac{\epsilon h}{\pi_1 {\cal I}_{KL}}) = o(1/h)$, as $h \rightarrow \infty$. 

It should be noted first that, similar to the standard CUSUM case, following \cite{ref_tar_book} (see p. 49), we have $\eta_n \rightarrow  \eta$ ($P_1$ almost sure), and $E_1[\eta_n] \rightarrow \bar \eta$ as $n \rightarrow \infty$, where 
$\bar \eta$ is a relatively small positive number compared to $\tilde S_n$, as $n \rightarrow \infty$. Therefore the additional assumption (ii) above follows 
easily. Assumption (i) on uniform integrability above follows from the fact that $E_1[Z_1^2]$ is finite (see Example 2.6.2 in \cite{ref_tar_book}).  Also, $\eta_n \geq 0$, and hence 
the assumption (iii) follows trivially. The main difficulty usually lies in verifying condition (iv). For the standard CUSUM algorithm, a sketch of a proof using a change of measure argument is provided in 
\cite{ref_tar_book} (see page 55, Example 2.6.2.). A similar argument can be used to prove the result for the current scenario. Considering that conditioned on a given time sequence of visits to state $1$ by the Markov chain $\xi_k$, the MRW considered here is a sum of i.i.d. random variables satisfying the same assumptions as in for the standard CUSUM case,  condition (iv) holds. Since this is true for all possible random sequences of times of visits to state $1$, the result holds by averaging over all possible such sequences as well. 
A more rigorous proof will be provided in a future extended version of this paper.

Next, we need a few notations borrowed from \cite{tartakovsky_fuh_HMM,lai_fuh_2001}. 
As before, define the first positive ladder epoch for the Markov random walk $\tilde S_n$ as $\tilde T_{+} = \inf\{n: \tilde S_n > 0\}$, and define the kernel 
$P_{+}(i, j, A) = P_1\{\xi_{\tilde T_{+}} = j, \tilde S_{\tilde T_{+}} \in A | \xi_0 = i\}$, $i, j \in \{0,1 \}$.  It can be then shown that under the existing assumptions for a strongly non-lattice MRW with a positive mean (as is the case here), the kernel $P_{+}$ is aperiodic and the associated ladder Markov chain $\xi_{\tilde T_{n^{+}}}$ has a stationary distribution $\pi_{+}$, where 
$\tilde T_{n^{+}}$ is the $n$-th ladder epoch of $\tilde S_n$. 
Finally, using another notation $\Delta(i), \; i \in \{0.1\}$, that is a solution to a Poisson equation (see (A.10) in \cite{tartakovsky_fuh_HMM}, further details omitted here due to space restrictions), we can state the following result regarding the expected first passage time $\hat \tau(h)$ adapting Proposition 3 (MNRT) from \cite{tartakovsky_fuh_HMM}:
\begin{align}
& E_1[\hat \tau(h)] = \frac{1}{\pi_1 {\cal I}_{KL}} \left( h + \frac{E_1 [ \tilde S_{T_{+}}^2]}{ E_1 [ \tilde S_{T_{+}}]} - \bar \eta \right. \nonumber \\
& \left. \:\:\:\: \:\:\:\: -\sum_{i=0}^1 \Delta(i)(\pi_{+}(i) - \mu(i)) \right) + o(1),
\label{NMRT_modified}
\end{align}
where $\mu(.)$ is the initial distribution of the Markov chain $\xi_k$. 

Noting that  one can choose the initial distribution $\mu$ to be the same as $\pi_{+}$ (although it is difficult to calculate), the last term inside the brackets in the above expression can be ignored and the following approximation can be used 
\begin{align}
& E_1[\hat \tau(h)] \approx  \frac{1}{\pi_1 {\cal I}_{KL}} \left( h + \frac{E_1 [ \tilde S_{T_{+}}^2]}{ E_1 [ \tilde S_{T_{+}}]} - \bar \eta \right),
\label{NMRT_approx}
\end{align}
which clearly resembles its counterpart for the case $E[H] \geq E_s$, given by \eqref{eq:LD_av_delay}. 

\subsection{Asymptotic Distribution of First Passage Time to a  False Alarm}
In this section, we consider the scenario where the MRW is operating under the no change hypothesis and denote the probability measure and expectations by $P_{\infty}, E_{\infty}$, respectively. We note that under $P_{\infty}$, the MRW 
$\tilde S_n$ has a negative drift $-\pi_1 I_0$.  In order to invoke the results  on limit distributions of maximal segmental scores of Markov-dependent partial sums from 
\cite{karlin_dembo_92},  we  assume that $Z_k$ takes both positive and negative values with positive probability. This is guaranteed when, for example, $f_1$ and $f_0$ are both Gaussian with different means etc.
It can also be shown that the matrix 
$$ \Phi(\gamma) = \left[ \begin{array}{cc} \tilde \alpha  & (1 - \tilde \alpha) E_{\infty}[\exp(\gamma Z_k)]  \\
                                                                   (1-\tilde \beta) & \tilde \beta E_{\infty}[\exp(\gamma Z_k)] \end{array} \right] $$
has a spectral radius $\rho(\gamma)$, which is log convex and $\rho(\gamma^*) = 1$ has a unique positive solution at
$\gamma^* =1$. We also assume $\tilde \alpha, \tilde \beta > 0$. 

It has been shown in \cite{karlin_dembo_92} that the asymptotic results for the run length to a false alarm for a MRW under the above conditions 
are independent of the initial state of the Markov chain $\xi_k$. We therefore fix the initial state $\xi_0 = 1$. 
We  define the negative ladder epochs $K_1, K_2, \ldots, K_n$ (with $K_0 = 0$), where 
$K_n = \inf \{k: k \geq K_{n-1}, \tilde S_k - \tilde S_{K_{n-1}} < 0\}, n = 1, 2, \ldots$. Clearly, $K_1$ is the first negative ladder epoch resulting in the reflected MRW $\bar W_{K_1} = 0$ for the first time after starting at $W_0=0$. Clearly,  $\bar W_{K_i} = 0, \; i=1, 2, \ldots, n$. In what follows, we will be interested in the tail probability of the maximum of $\bar W_k$ in each of these positive excursions between $K_{i-1} \leq k \leq K_i$, and eventually the tail probability of the maximum of all these maximums. Similar to \cite{seq_det_khan, karlin_dembo_92}, it can be shown that the tail probability of the first passage time to a false alarm $P_{\infty}(\hat \tau_{\infty}(h) > n)$ is the same as the probability 
$P_{\infty}(M_n < h)$, where $M_n = \max \{Q_1, Q_2, \ldots, Q_{R_n}, Q^*\}$, $Q_i$ is the maximum of the reflected 
MRW during the $i$-the positive excursion, $R_n$ is the number of such positive excursions before time $n$, and $Q^*$ is the maximal segmental score between time $K_{R_n}$ and $n$. Note also that since the MRW $\tilde S_n$ only increments when the Markov chain $\xi_k$ visits state $1$, the states the (negative) ladder Markov chain visits at 
times $K_1, K_2, \ldots, $ are also $1$. This implies that each nonnegative excursion of  the Markov chain begins and ends at state $1$ only, and therefore the ladder Markov chain only has a single state $1$. This simplifies the calculations significantly. Note also that the maximum of the individual excursion period $Q_i$ is independent and identically distributed, and since within each excursion the MRW is a sum of i.i.d. random variables  $\tilde S_n  = \sum_{k=0}^n \tilde S_{t_k}$, where $0 = t_0 < t_1 < \ldots < t_n \leq n$, 
applying Equation (2.10) from \cite{seq_det_khan}, and simplifying the analysis for the MRW case from \cite{karlin_dembo_92}, one can show that the asymptotic tail distribution of the maximum of the first non-zero excursion in the MRW case is the same as that in the i.i.d. case, that is, 
\begin{align}
& \lim_{h \rightarrow \infty} P_{\infty}\left(e^h P(Q_1 > h\right) =  c(\infty) =\frac{(1-E_{\infty} [e^{ S_{T_{-}}}])^2}{{\cal I}_{KL}E_{\infty}[T_{-}]},
\label{eq:dist_max1}
\end{align}
where $S_{T_{-}}, T_{-}$ are defined as the first negative ladder height and the the first negative ladder epoch for the regular random walk with i.i.d. increments discussed in 
Section III.B.  Further technical details of this result will be provided in an extended version of this work.

Finally, invoking Theorem B from \cite{karlin_dembo_92} (see p. 118), and simplifying to the current scenario, we can state the following result:
\begin{thm}
For an energy harvesting sensor employing  a CUSUM test \eqref{cusum_mod}  with an average harvested energy $\bar H < E_s$, the asymptotic tail distribution of the (normalized) first passage time to a false alarm  is given by 
\begin{align}
&  P_{\infty} (e^{-h} \hat \tau_{\infty}(h) > x) = e^{- \beta_{MRW} x}, \; h \rightarrow \infty,
\label{tail_dist_FA_MRW}
\end{align}
where $ \beta_{MRW} =-\frac{\pi_1 I_0 c(\infty)}{E_{\infty}[\tilde S_{K_1}]}$, $c(\infty)$ is given by \eqref{eq:dist_max1}, and $\pi_1$ is the stationary probability of the underlying Markov chain $\xi_k$ being in state $1$.  Similarly, $E_{\infty}[\hat \tau_{\infty}(h)] = \frac{e^h}{\beta_{MRW}}$. 
\end{thm}
\begin{remark}
Note that the negative sign in at the front of the expression is due to the fact that the mean of the MRW is $\pi_1 I_0$, but note also that $E_{\infty}[\tilde S_{K_1}]$ is negative, therefore $ \beta_{MRW}$ is positive.
\end{remark}

\section{Numerical Results}
In this section, we provide some numerical results, where an energy harvesting sensor is employed to detect a change in mean of a Gaussian distribution ${\cal N}(0, \sigma^2)$ to 
${\cal N}(m_1, \sigma^2)$, where  $m_1=0.5$, $ \sigma^2 = 1$.  $E_s$ is chosen as $0.5$ milli Joule (mJ). We run Monte Carlo simulations over $30000$ samples and average over $150000$ simulation runs to obtain the following results regarding the expected detection delay and the exponent of the asymptotically exponential tail distribution for the first passage time to a false alarm for both $\bar H \geq E_s$ and $\bar H < E_s$.  The threshold for detection is $h = 10$. We also note that ${\cal I}_{KL} = - I_0 = \frac{m_1^2}{2\sigma^2}$, and the change occurs at $\nu=1$. 

Table \ref{tab:table1} below shows the expected detection delay computed theoretically (based on \eqref{eq:LD_av_delay} or \eqref{NMRT_approx}) and the corresponding value obtained through simulations for different values of $\bar H \geq E_s$ and also $\bar H < E_s$. The corresponding average run legth to false alarm can be approximated 
as $\frac{e^h}{\bar \beta}$ for $\bar H \geq E_s$ or $\frac{e^h}{\beta_{MRW}}$ for $\bar H < E_s$. \\

\begin{table}[h!]
  \begin{center}
    \caption{Expected detection delay (in number of samples)}
    \label{tab:table1}
    \begin{tabular}{l|S|r} 
      \textbf{$\bar H (mJ)$} & \textbf{Theoretical} & \textbf{Simulated}\\
      \hline
      0.7 & 76.5907  & 76.6696\\
      0.6 & 76.6850  & 76.6481 \\
      0.5  &76.6787  & 76.7750\\
      0.4 &  95.7702  & 95.9735  \\
      0.3 &   127.8217  &  127.2639\\
      0.2 &    191.6132 &  189.2639 
    \end{tabular}
  \end{center}
\end{table}
The values of $\bar \beta$ obtained from simulations (when $\bar H \geq E_s$) is $0.0699$, whereas the corresponding $\beta_{MRW}$ values for $\bar H < E_s$ are computed as 
$0.0558, 0.0417, 0.0283$ for $\bar H = 0.4, 0.3, 0.2$ mJ, respectively. Figure \ref{fig:fig1} below shows that the tail probability exponent for $\bar H =0.4$ asymptotically approaching close to the theoretically calculated value $\beta_{MRW} = 0.0558$.
\begin{figure}[h!]
\centering
\includegraphics[scale=0.40]{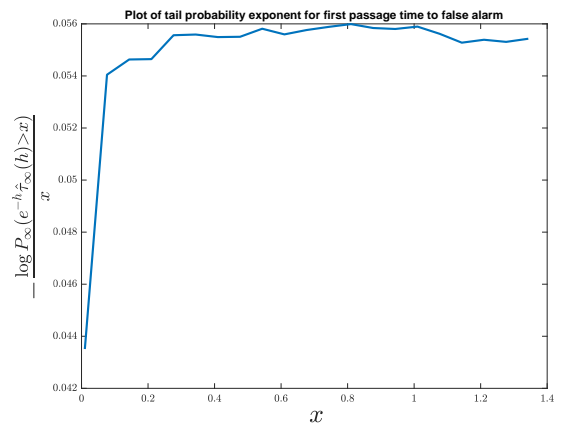}
\caption{Tail probability exponent for first passage time to a false alarm for $\bar H = 0.4$,  $h=10$}
\label{fig:fig1}
\end{figure}
\section{Conclusions}
In this paper, we presented asymptotic results regarding the expected detection delay and the tail distribution of the run-length to a false alarm when an energy harvesting sensor is employed to perform a sequential change detection task using the CUSUM method. It is seen that the analysis can be divided into two distinct scenarios, (i) $\bar H \geq E_s$, and (ii) $\bar H < E_s$. While standard existing asymptotic results for the CUSUM test apply in the first case, the second scenario is more complicated and requires asymptotic results from Markov random walks and associated nonlinear Markov renewal theory. Future work will consider decentralized sequential change detection with multiple sensors employing local detection and a fusion centre implementing a global decision.
\bibliographystyle{IEEEtran}
\renewcommand{\baselinestretch}{0.9}
\setlength{\itemsep}{-0.1mm}
%\bibliography{IEEEabrv,Quick_ref}

\begin{thebibliography}{10}
\providecommand{\url}[1]{#1}
\csname url@samestyle\endcsname
\providecommand{\newblock}{\relax}
\providecommand{\bibinfo}[2]{#2}
\providecommand{\BIBentrySTDinterwordspacing}{\spaceskip=0pt\relax}
\providecommand{\BIBentryALTinterwordstretchfactor}{4}
\providecommand{\BIBentryALTinterwordspacing}{\spaceskip=\fontdimen2\font plus
\BIBentryALTinterwordstretchfactor\fontdimen3\font minus
  \fontdimen4\font\relax}
\providecommand{\BIBforeignlanguage}[2]{{%
\expandafter\ifx\csname l@#1\endcsname\relax
\typeout{** WARNING: IEEEtran.bst: No hyphenation pattern has been}%
\typeout{** loaded for the language `#1'. Using the pattern for}%
\typeout{** the default language instead.}%
\else
\language=\csname l@#1\endcsname
\fi
#2}}
\providecommand{\BIBdecl}{\relax}
\BIBdecl


\bibitem{ref15}
H.~Poor and O.~Hadjiliadis, \emph{Quickest Detection}.\hskip 1em plus 0.5em
  minus 0.4em\relax Cambridge University Press, 2008.

%\bibitemG.~Lorden, ``Procedures for reacting to a change in distribution,'' \emph{Ann.
%  Math. Statist.}, vol.~42, no.~6, pp. 1897--1908, 12 1971.

\bibitem{ref19}
M.~Pollak, ``Optimal detection of a change in distribution,'' \emph{Ann.
  Statist.}, vol.~13, no.~1, pp. 206--227, 03 1985.
\bibitem{ref16}
A.~G. Tartakovsky and V.~V. Veeravalli, ``Asymptotically optimal quickest
  change detection in distributed sensor systems,'' \emph{Sequential Analysis},
  vol.~27, no.~4, pp. 441--475, 2008.

\bibitem{ref20}
J.~Geng and L.~Lai, ``Non-bayesian quickest change detection with stochastic
  sample right constraints,'' \emph{IEEE Transactions on Signal Processing},
  vol.~61, no.~20, pp. 5090--5102, Oct 2013.

\bibitem{ref21}
J.~Geng, E.~Bayraktar, and L.~Lai, ``Bayesian quickest change-point detection
  with sampling right constraints,'' \emph{IEEE Transactions on Information
  Theory}, vol.~60, no.~10, pp. 6474--6490, Oct 2014.


\bibitem{ref_tar_book}
A.G. Tartakovsky, I.~Nikiforov, and M.~Basseville, \emph{Sequential Analysis: Hypothesis
  Testing and Change Point Detection}.\hskip 1em plus 0.5em minus 0.4em\relax
  Boca Raton, FL, USA: CRC Press, Taylor and Francis Group, 2015.

\bibitem{iglehart}
D.~L.~Iglehart, ``Extreme Values in the {GI/G/1} Queue,'' {\em The Annals of Mathematical Statistics}, vol.~43, n0.~2, pp.~627-635, 1972.

\bibitem{seq_det_khan}
R.~A.~Khan, ``Detecting changes in probabilities of a multi-component process,'' {\em Sequential Analysis}, vol.~14, no.~4, pp.~375-388, 1995.
\bibitem{siegmund}
D.~Siegmund, {\em Sequential Analysis: Tests and Confidence Intervals}, Springer Series in Statistics, Springer-Verlag, New York, 1985. 

\bibitem{pollak_tar_09}
M.~Pollak and A.~G.~Tartakovsky, ``Asymptotic Exponentiality of the Distribution of First Exit Times  for a Class of Markov Processes with Applications to 
Quickest Change Detection,'' {\em Theory of Probability and Its Applications {(SIAM)}}, vol.~53, no.~3, pp.~430-442,  Aug. 2009.

\bibitem{meyn_tweedie_book}
S.~Meyn and R.~L.~Tweedie, {\em Markov Chains and Stochastic Stability}, 2nd Edition, Cambridge University Press, Cambridge, UK, 2009.

\bibitem{lai_fuh_98}
C-D. Fuh and T.~Z. Lai, ``Wald's Equations, First Passage Times and Moments of Ladder Variables in Markov Random Walk,'' {\em Journal of Applied Probability}, 
vol. 35, no. 3, pp.~566-580, Sept. 1998.

\bibitem{tartakovsky_fuh_HMM}
C-D. Fuh and A.~G. Tartakovsky, ``Asymptotic Bayesian Theory of Quickest Change Detection for Hidden Markov Models,'' 
{\em IEEE Transactions on Information Theory}, vol. 65, no. 1, pp.~511-529, Jan. 2019.

\bibitem{lai_fuh_2001}
C-D. Fuh and T.~Z. Lai,  ``Asymptotic Expansions in Multidimensional Markov Renewal Theory and First Passage Times for Markov Random Walks,'' 
{\em Advances in Applied Probability}, vol.~33, no.~3, pp.~652-673, Sep. 2001.

\bibitem{karlin_dembo_92}
S.~Karlin and A.~Dembo, ``Limit Distributions of Maximal Segmental Score among Markov-Dependent Partial Sums,'' {\em Advances in Applied Probability},
vol.~24, no.~1, pp.~113-140, Mar. 1992.

\end{thebibliography}

\end{document}